\documentclass[useAMS,usenatbib]{mn2e}
\usepackage{graphicx}
\title[SOAR-OSIRIS near-infrared spectra of a new Galactic O2If* star]{An O2If* star found in isolation in the backyard of 
NGC 3603\thanks{Based on observations obtained at the Southern 
Astrophysical Research (SOAR) telescope}}
\author[A. Roman-Lopes]{A. Roman-Lopes$^{1}$\thanks{roman@dfuls.cl}\\
$^{1}$Department of Physics - Universidad de La Serena - Cisternas, 1200 - La Serena - Chile\\
}

\begin{document}

\date{}

\pagerange{\pageref{firstpage}--\pageref{lastpage}} \pubyear{2010}

\maketitle

\label{firstpage}

\begin{abstract}

In this letter we communicate the identification of a new Galactic O2If* star (MTT 68) isolated at a projected linear distance of 3 pc from the centre of 
the star-burst cluster NGC 3603. From its optical photometry I computed a bolometric luminosity
M$_{Bol}$ = -10.7, which corresponds to a total stellar luminosity of 1.5$\times$10$^{6}$ L$_\odot$.
It was found an interesting similarity between MTT 68 and the well known multiple system HD 93129. 
From Hubble Space Telescope F656N images of the NGC 3603 field, it was found that MTT 68 is actually a visual binary system with an angular 
separation of 0.38$\arcsec$, which 
corresponds to a projected (minimum) linear distance of \textit{r}$_{A-B}$ = 1.4$\times$10$^{-2}$ pc. This value is similar to that for the HD 93129A (O2If*) and 
HD 93129B (O3.5) pair, \textit{r}$_{A-B}$ = 3.0$\times$10$^{-2}$ pc. On the other hand, HD93129A has a third closer companion named HD 93129Ab (O3.5) at only 
0.053$\arcsec$, and taking into account that the
X-ray to total stellar luminosity ratio for the MTT 68 system (L$_X$/L$_{Bol}$ $\sim$1$\times$10$^{-5}$) is about two orders of
magnitude above the canonical value expected for single stars, I suspect that the MTT 68 system probably hosts another massive companion possibly
to close to be properly resolved by the HST archive images.

\end{abstract}


\section{Introduction}

NGC 3603 is the closest star-burst like cluster, being an invaluable template for the modern theory of formation and evolution of \textit{very massive stars}. 
Indeed, with at least four exemplars in its core (three WN6ha + one O2If*/WN6 - \citet{b7}), plus two O2If*/WN6 stars (WR42e and MTT58) recently identified at only a 
few arcminutes from its centre \citep{b8,b9}, 
this cluster may be the host of the larger concentration of extremely massive \textit{hydrogen core burning} stars in the Galaxy \citep{b10,b3,b7}.

\citet{b6} detected a large number of X-ray sources toward the cluster centre, concluding that the majority of them should be probably compound by pre-main
sequence stars. However, the detection of two O2If*/WN6 stars in the NGC 3603's periphery may indicate that some other very massive stars could be present in 
the NGC 3603 cluster field. In this context and based on the presence of very strong X-Ray counterparts in the BMW-Chandra catalogue \citep{b13}, 
we performed NIR 
follow-up spectroscopic observations of a very interesting X-ray point source previously cataloged by \citet{b2} as MTT 68. It is now confirmed to be an O2If* star 
isolated in the periphery of NGC 3603, at about 1.4$\arcmin$ from its core.

\section{Near-Infrared spectroscopic observations and data reduction}

The NIR spectroscopic observations were performed with the Ohio State Infrared Imager and Spectrometer (OSIRIS)
at the Southern Astrophysics Research (SOAR) telescope. The J-, H- and K-band data were acquired in 31th January 2012 with the night presenting good
atmospheric conditions.

In Table 1 it is shown a summary of the NIR observations. 
The raw frames were reduced following well known NIR reduction procedures. 
The two-dimensional frames were subtracted for each pair of images taken
at the two shifted positions, with the resultant images being divided
by a master normalized flat. For each processed frame, the J-, H- and K-band spectra were extracted using the task
APALL within IRAF, with the wavelength calibration made through the IRAF tasks IDENTIFY and DISPCOR applied to a set of OH sky line spectra in the 
range 12400\AA\ -23000\AA\ . 
The corrections of the telluric atmospheric features on the science data, were performed using J-, H- and K-band spectra of A
type stars obtained before and after the science targets. 
The photospheric absorption lines present in the high signal-to-noise telluric
spectra, were subtracted from a careful modeling (through the use of Voigt and
Lorentz profiles) of the hydrogen absorption lines using the corresponding adjacent continuum. 
At the end, the J-, H- and K-band spectra were combined by using the IRAF task SCOMBINE with the mean signal-to noise ratio of the resulting spectra
well above 100.


\begin{table}
\caption{Summary of the SOAR/OSIRIS spectroscopic observations.}
\label{catalog}
\centering
\renewcommand{\footnoterule}{}  
\begin{tabular}{cc}
\hline
   Date  & 31/01/2012\\
   Telescope  & SOAR\\
   Instrument & OSIRIS\\
   Mode  & XD\\
   Camera & f/3\\
   Slit  & 1" x 27"\\
   Resolution  & 1000\\
   Coverage ($\mu$m) & 1.25-2.35\\
   Seeing (")  & 0.8-1.3\\
\hline
\end{tabular}
\end{table} 

   \begin{figure}
    \vspace{0pt}
    \hspace{-5pt}
   \centering
   \includegraphics[bb=14 14 512 337,width=8.7 cm,clip]{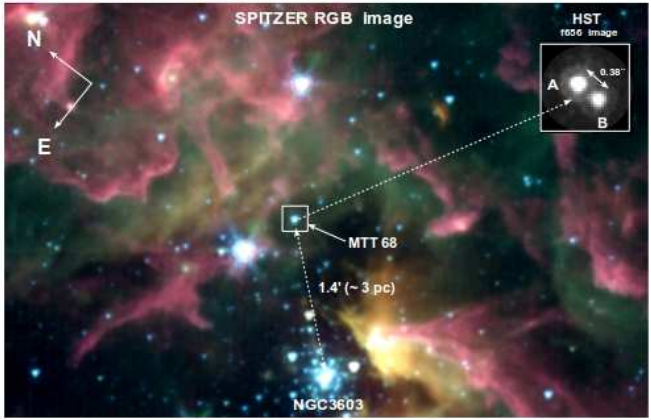}
     \caption{The three color image of the north and west regions towards the NGC 3603 cluster's centre, constructed from the 3.6$\mu$m 
     (blue), 4.5$\mu$m (green) and 
     5.8$\mu$m (red) Spitzer IRAC data taken from the NASA/IPAC Infrared Science Archive. Notice that the new star is found in isolation at about 1.4 arcmin 
     from the cluster's centre, which for an estimated heliocentric distance of 7.6 kpc corresponds to a projected distance of about 3 parsecs.}
         \label{FigVibStab}
   \end{figure}

   \begin{figure}
    \vspace{0pt}
    \hspace{0pt}
   \centering
   \includegraphics[bb=14 14 423 296,width=8.7 cm,clip]{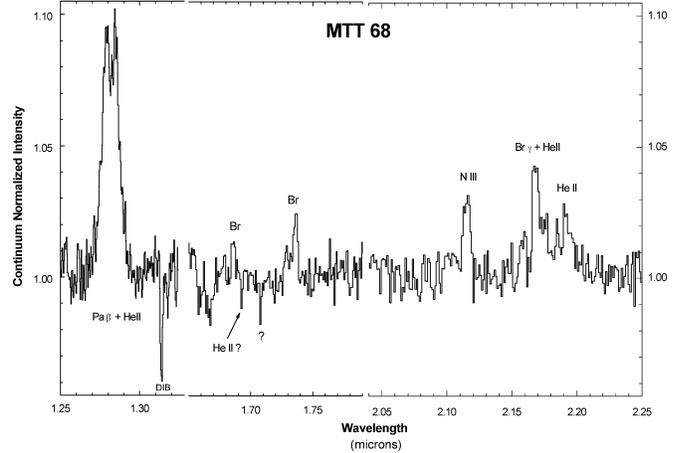}
     \caption{The J- H- and K-band continuum normalized SOAR-OSIRIS spectra of MTT 68, with the main H, He and N emission lines identified by labels.}
         \label{FigVibStab}
   \end{figure} 

\section{Results}

Coordinates and photometric parameters of MTT 68 are shown in Table 2. The V-, R- and I-band optical magnitudes were taken from
the work of \citet{b20}, while the near-infrared values were obtained from the Two-Micron All Sky Survey \citep{b21}, with the absorption-corrected 0.5-10keV
Chandra X-ray flux taken from the work of \citet{b13}.

\subsection{The OSIRIS NIR spectra of MTT 68: An O2If* star isolated in the backyard of NGC 3603}

As mentioned before, the star subject of this letter was previously cataloged as MTT 68 by \citet{b2}. In Figure 1 it is shown a three colour image of part 
of the NGC 3603 field, made from the 3.6$\mu$m (blue), 4.5$\mu$m (green) and 5.8$\mu$m (red) Spitzer 
IRAC data taken from the NASA/IPAC Infrared Science Archive\footnote{http://irsa.ipac.caltech.edu/data/SPITZER/docs/spitzerdataarchives/}. From that figure, we 
can see that 
the new star appears isolated at about 1.4 arcmin from the cluster's centre, which for an estimated heliocentric distance of 7.6 kpc \citep{b3} 
corresponds to a projected radial distance of about 3 parsecs. From a search in the CADC HST/HLA/WFPC2B Science 
Archive\footnote{http://www3.cadc-ccda.hia-iha.nrc-cnrc.gc.ca/hst/new/} for optical images of the NGC 3603 field, it was found a non-saturated F656N image 
(IB6WA1070 - P.I. O'Connell) in which we 
see that MTT 68 has a visual companion (MTT 68B - please see the inset in Figure 1) at an angular separation of 0.38$\arcsec$. More on this subject will be 
discussed in Section 3.2.

Figure 2 shows the telluric corrected (continuum normalized) J-, H- and K-band SOAR-OSIRIS spectra of MTT 68, in which we can see the Paschen beta and Bracket
hydrogen emission lines at 1.283$\mu$m, 1.736$\mu$m and 2.166$\mu$m, as well as the N{\sc iii} and He{\sc ii} lines (also in emission) at 2.116$\mu$m and 2.189$\mu$m,
respectively. 
In Figure 3 it is presented the MTT 68's J-, H- and
K-band SOAR-OSIRIS spectra, together with those taken for HD93129A (O2If*), WR20a (O3If*/WN6 + O) 
and WR25 (O2.5If*/WN6). The spectral type O2If* was introduced in 2002 \citep{b32}, with HD93129A being probably the only known Galactic template of the class. It is considered the earliest, hottest, 
most massive and luminous O star in the Galaxy, showing an extremely powerful wind with a terminal velocity above 3000 km s$^{-1}$, and a mass-loss rate above 
10$^{-5}$ M$_{\odot}$ yr$^{-1}$ \citep{b31}.
From the comparison of the MTT 68's NIR spectra with those for the templates, we can see that
the MTT 68's spectrograms resembles well those for HD93129A, indicating that it is probably a new Galactic exemplar of the rare O2If* type. 

   \begin{figure*}
   \vspace{0pt}%
   \hspace{-10pt}
  \centering
  \includegraphics[bb=14 14 671 508,width=10 cm,clip]{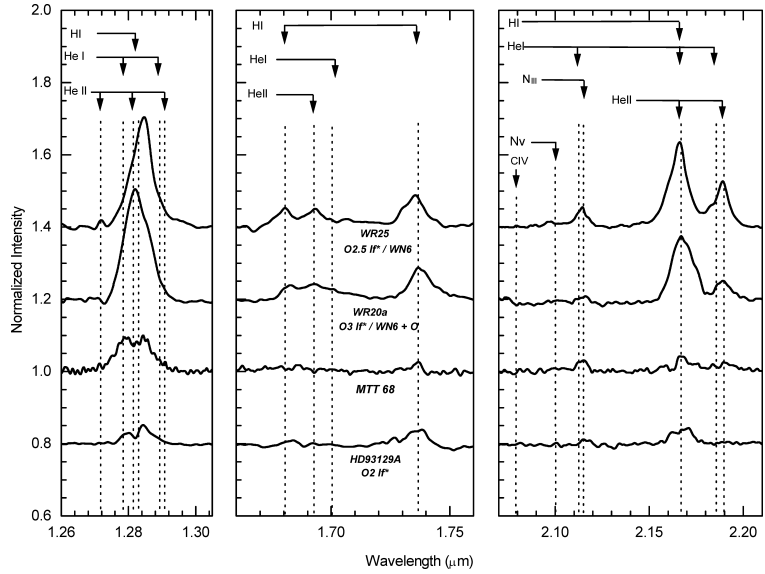}
     \caption{The J- H- and K-band continuum normalized SOAR-OSIRIS spectra of the MTT 68, together with the NIR spectrograms of
               HD93129A (O2If*), WR20a (O3If*/WN6 + O3If*/WN6) and WR25 (O2.5If*/WN6), with the main H, He and N emission lines identified by labels.
               By comparing the MTT 68's NIR spectra with those for the templates, we can see that the MTT 68's spectrograms resembles well those of HD93129A, 
               indicating that it is probably a new Galactic exemplar of the O2If* type.}
        \label{FigVibStab}
   \end{figure*}

   \begin{figure*}
   \vspace{0pt}%
   \hspace{-10pt}
  \centering
  \includegraphics[bb=14 14 326 434,width=5 cm,clip]{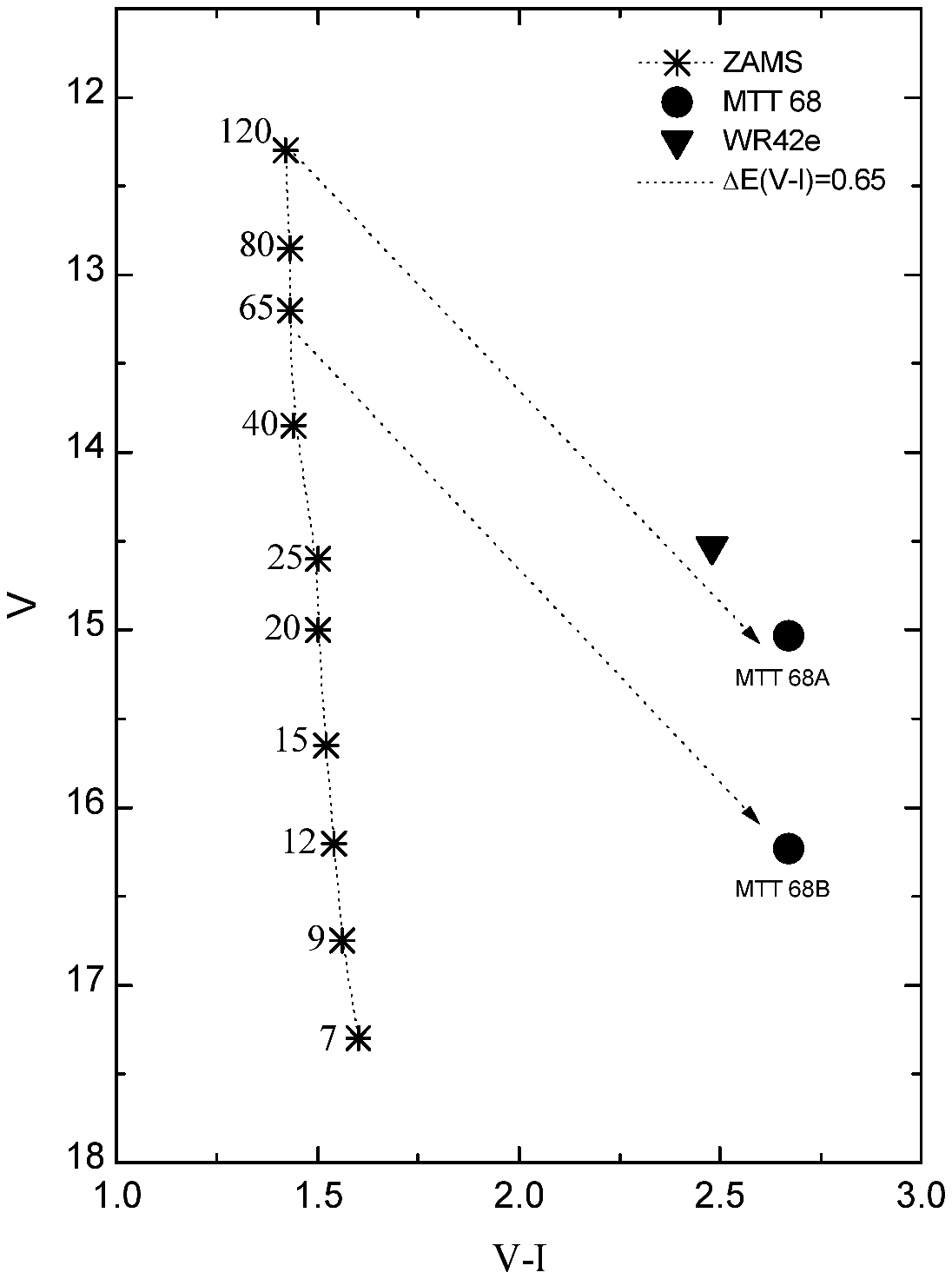}
     \caption{The V $\times$ (V-I) diagram (based on Figure 7 of the work of \citet{b20}, with MTT 68A and MTT 68B represented by the black circles, 
     and WR42e $\sim$ 130M$_\odot$ - \citep{b24} by the black triangle. The non-reddened main sequence at the quoted distance of 7.6 kpc \citep{b3} 
     for masses between 7M$_\odot$ to 120M$_\odot$ is represented by the black dotted vertical line, with the black stars indicating the position of each mass bin. Also, 
     the reddening vector taken from the work of \citet{b20} is represented by the line-dotted arrows. From this diagram we can see that the initial mass of MTT 68A 
     (the new O2If* star), possibly exceed 100M$_\odot$. On the other hand, despite the fact that MTT 68B is probably 1.2 magnitudes fainter than MTT 68A, it is still
     a very massive star with probable initial mass well above 40M$_\odot$.}
        \label{FigVibStab}
   \end{figure*}   
 



\begin{table*}
\begin{center}
\caption{Coordinates (J2000), Optical/NIR photometry, and X-ray parameters of MTT 68.
The BVRI photometry was taken from \citet{b20}, while the near-infrared magnitudes are from \citet{b21}, and the absorption-corrected 0.5-10keV flux was obtained from the 
work of \citet{b13}.\label{tbl-2}}
\begin{tabular}{ccccccccccc}
\\
\hline\hline
RA  &Dec &B &V &R &I &J &H &Ks &X-Ray& \\
(J2000) & (J2000) & & & & & & & & (Wm$^{-2}$)\\
\hline
11:14:59.48 &-61:14:33.9 &16.31 &14.72 &13.64 &12.05 &9.98 &9.17 &8.74 &12.9$\times$10$^{-16}$& \\
\hline\hline

\end{tabular}
\end{center}
\end{table*}

\subsection{Mass, luminosity and binary status of MTT 68}

We can estimate the mass of the MTT 68 components taking into account that the star is actually a visual binary system presenting an angular 
separation of 0.38$\arcsec$, and by comparing their combined V-band magnitude and V-I color 
with those of other NGC 3603 cluster members, presented in Figure 7 of \citet{b20}. Also, from 
the comparison of the instrumental magnitudes of the two (non-saturated) point sources in the HST F656N image, 
we estimate as 1.2 mag the $\Delta_m$ difference for the MTT 68A and MTT 68B. Finally, and in order to
simplify the process, we will assume that both stars have the same bolometric corrections, which is a reasonable assumption considering 
their probable very early spectral types.
In Figure 4 we show an adapted version of the V $\times$ (V-I) diagram for NGC 3603 \citep{b20}, with MTT 68A, MTT 68B and WR42e (estimated mass 
of $\sim$ 130 M$_\odot$ - \citet{b8,b24} represented by black circles and gray triangle, respectively. WR42e is also shown there
because it is a star of the O2If*/WN6 spectral type and as the MTT 68 system, possibly belongs to the same complex.
From this diagram we can see that MTT 68A and MTT 68B probably presented initial masses  
well above 100 M$_\odot$ and 40 M$_\odot$, respectively. 

To estimate the total luminosity of the MTT 68 binary system, it is necessary to compute the associated visual extinction considering that 
the interstellar reddening law for NGC 3603 is possibly abnormal, with a ratio of total to selective extinction value R$_V$=3.55$\pm$0.12 \citep{b20}.
From Table 2, we can see that
MTT 68 presents (B-V) color $\sim$ 1.6 mag, which for an assumed mean intrinsic (B-V)$_0$ value of -0.3 mag (typical for the hottest early-type stars), corresponds 
to a color excess E(B-V) $\sim$ 1.9 mag or A$_V$ $\sim$ 6.7$\pm$0.3 mag. 
From the computed color excess and visual extinction, we can estimate the MTT 68's absolute magnitude using the distance modulus equation, 
assuming that the star is part of the NGC 3603 complex at an heliocentric distance of 7.6$\pm$0.4 kpc \citep{b3}. We computed M$_V$=-6.4 mag for the binary system
(or individually M$_V$=-6.1 and M$_V$=-4.9 for components \textit{A} and \textit{B}, respectively), 
which for an assumed mean bolometric correction BC $\sim$ -4.3 mag \citep{b3,b7}, results in a bolometric magnitude 
M$_{Bol}$ $\sim$ -10.7 and in a total stellar luminosity of 1.5$\times$10$^{6}$ L$_\odot$. This total luminosity is similar to that derived by \citet{b6}, which
found L$\sim$1.3$\times$10$^{6}$ L$_\odot$ with an X-ray to total stellar luminosity ratio L$_X$/L$_{Bol}$ $\sim$1$\times$10$^{-5}$, a value two orders of
magnitude greater than the canonical value expected for single stars, e.g. L$_X$/L$_{Bol}$ $\sim$ 10$^{-7}$ \citep{b31}.

\subsection{Similarities with the HD93129 system}

From the measured angular separation (0.38$\arcsec$) and the assumed heliocentric distance of 7.6 kpc, it is possible to compute the linear 
projected (minimum) radial distance of the MTT 68 binary components as \textit{r}$_{A-B}$ = 1.4$\times$10$^{-2}$ pc. HD93129A is known to be an extremely powerful X-ray 
source, being part of the Trumpler 14 cluster in the Carina
Nebula, at an heliocentric distance of 2.3 kpc (for more on it please see the work of \citet{b31} and references therein). As the new O2If* star, it has a 
visual companion named HD93129B (O3.5) at an angular separation of 2.7$\arcsec$, which for the
quoted distance corresponds to a linear projected radial distance of 3$\times$10$^{-2}$ pc, a value similar to that for the MTT 68 components. On the other hand,
HD93129A has also another closer companion (HD 93129Ab) at only 0.053$\arcsec$ that is supposed to be also of the same spectral type of HD 93129B \citep{b31}. Considering the observed
X-ray to total stellar luminosity ratio of the MTT 68 system (L$_X$/L$_{Bol}$ $\sim$1$\times$10$^{-5}$), it is reasonable to speculate that this might be also the case for 
MTT 68, e.g., the probable existence
of a very close companion not resolved in the F656N HST images. 

Certainly, new further spectro-photometric studies of the MTT 68 binary system are needed. In one hand, 
its present evolutionary stage can help us to better understand how such kind of system can be found (and build) in relative isolation 
(at about 3.1 pc from the NGC 3603 cluster centre). On the other hand (and may be even more important), its presence in the periphery of a star burst like cluster
may represent a challenge to the present theory of formation and evolution of very massive stellar systems, in the sense that may be the unique conditions found in such kind
of environment could produce very massive stars well beyond the core of stellar clusters like NGC 3603. 

\section{Summary}

In this work we communicate the identification of a new Galactic O2If* star (MTT 68) that is found in isolation at about 1.4$\arcmin$ (about 3 pc considering a 
quoted heliocentric distance of 7.6 kpc) from the core of the star-burst cluster NGC 3603. Our main conclusions and results are:

1- By comparing its NIR spectra with those for HD 93129A (O2If*), WR20a (O3If*/WN6), and WR25 (O2.5If*/WN6), we conclude that
the MTT 68's spectrograms resembles well those of the former, indicating that it is probably a new Galactic exemplar of the rare O2If* type. 

2- From the inspection of F656N HST images of NGC 3603, it was determined that MTT 68 is actually a visual binary presenting an angular separation of 0.38$\arcsec$.
Also, from the instrumental photometry of the non-saturated MTT 68 point sources, it was possible to determine a magnitude difference $\Delta_m$ = 1.2 magnitudes,
which combined with the B-, V- and I-band photometry taken from the literature, resulted in a combined bolometric luminosity
M$_{Bol}$ = -10.7 (M$_{Bol}$ = -10.4 and -9.2 for MTT 68A and MTT 68B, respectively) equivalent to a total stellar combined luminosity of 1.5$\times$10$^{6}$ L$_\odot$.

3- The total luminosity derived by us is similar to that obtained by \citet{b6}, which
found L$\sim$1.3$\times$10$^{6}$ L$_\odot$ with an X-ray to total stellar luminosity ratio L$_X$/L$_{Bol}$ $\sim$1$\times$10$^{-5}$, a value two orders of
magnitude above the canonical value for single stars, e.g. L$_X$/L$_{Bol}$ $\sim$ 10$^{-7}$ \citep{b31}.

4- From the associated V- and I-band photometry, the magnitude differences obtained from the HST F656N image and stellar models previously published for the NGC 3603 
massive stellar population, it was possible to conclude that MTT 68A and MTT 68B probably presented initial masses above 100 M$_\odot$ and 40 M$_\odot$, respectively. 

5- We found some interesting similarities with the well known multiple system HD 93129. In one hand, the measured angular separation of the MTT 68 binary components 
(0.38$\arcsec$) corresponds to a projected (minimum) linear distance of \textit{r}$_{A-B}$ = 1.4$\times$10$^{-2}$ pc, a value similar to that for HD 93129A (O2If*) 
and HD 93129B (O3.5), e.g.,
\textit{r}$_{A-B}$ = 3.0$\times$10$^{-2}$ pc. On the other hand, HD93129A has another closer companion named HD 93129Ab (O3.5) at only 0.053$\arcsec$ \citep{b31}. 
Considering the observed
X-ray to total stellar luminosity ratio for the MTT 68 system (L$_X$/L$_{Bol}$ $\sim$1$\times$10$^{-5}$), it is possible that some another very close massive companion, 
not resolved in the F656N HST images, is still to be detected in the MTT 68's binary system.

\section*{Acknowledgments} 

I would like to thank the anonymous referee by the careful reading of the manuscript. Her/his
comments and criticism were welcome.
This research has made use of the NASA/ IPAC Infrared Science Archive, which
is operated by the Jet Propulsion Laboratory, California Institute of
Technology, under contract with the National Aeronautics and Space
Administration. 
This publication makes use of data products from the Two Micron All Sky
Survey, which is a joint project of the University of Massachusetts and the
Infrared Processing and Analysis Center/California Institute of Technology,
funded by the National Aeronautics and Space Administration and the National
Science Foundation. 
Based on observations obtained at the Southern Astrophysical Research (SOAR) telescope, which is a joint project of the 
Minist\'{e}rio da Ci\^{e}ncia, Tecnologia, e Inova\c{c}\~{a}o (MCTI) da Rep\'{u}blica Federativa do Brasil, the U.S. National 
Optical Astronomy Observatory (NOAO), the University of North Carolina at Chapel Hill (UNC), and Michigan State University (MSU).
Based (in part) on observations made with the NASA/ESA Hubble Space Telescope, and obtained from the Hubble Legacy Archive, which is a collaboration 
between the Space Telescope Science Institute (STScI/NASA), the Space Telescope European Coordinating Facility (ST-ECF/ESA) and the Canadian Astronomy 
Data Centre (CADC/NRC/CSA).
This work was partially supported by the Department of Physics of the Universidad de La Serena.
ARL thanks financial support from Diretoria de Investigaci\'on - Universidad de La Serena through Project \textquotedblleft 
DIULS REGULAR PR13144\textquotedblright.

\end{document}